\documentclass[apjl]{emulateapj}

\newcommand{\be}{\begin{equation}}
\newcommand{\ee}{\end{equation}}
\newcommand{\astrosat}{{\it AstroSat }}
\newcommand{\source}{Mrk 421}

\begin{document}

\title{Possible Accretion Disk Origin of the Emission Variability of a Blazar Jet}

\author{Ritaban Chatterjee\altaffilmark{1,*}, Agniva Roychowdhury\altaffilmark{1}, Sunil Chandra\altaffilmark{2,3}, Atreyee Sinha\altaffilmark{4,5}}

\altaffiltext{1}{Department of Physics,  Presidency University, 86/1 College Street, Kolkata 700073.}
\altaffiltext{2}{Tata Institute of Fundamental Research, Mumbai, India, 400005.}
\altaffiltext{3}{Centre for Space Research, North-West University, Potchefstroom, 2520, South Africa.}
\altaffiltext{4}{Inter University Center for Astronomy and Astrophysics, Pune - 411 007, India.}
\altaffiltext{5}{AstroParticule et Cosmologie, Universite Paris Diderot, CNRS/IN2P3, Paris 75013.}
\altaffiltext{*}{ritaban.physics@presiuniv.ac.in}

\begin{abstract}
We analyze X-ray light curves of the blazar Mrk 421 obtained from the Soft X-ray Imaging Telescope and the Large Area X-Ray Proportional Counter instrument onboard the Indian space telescope \textit{AstroSat} and archival observations from \textit{Swift}. We show that the X-ray power spectral density (PSD) is a piece-wise power-law with a break, i.e., the index becomes more negative below a characteristic ``break-timescale". Galactic black hole X-ray binaries and Seyfert galaxies exhibit a similar characteristic timescale in their X-ray variability that is proportional to their respective black hole mass. X-rays in these objects are produced in the accretion disk or corona. Hence, such a timescale is believed to be linked to the properties of the accretion flow. Any relation observed between events in the accretion disk and those in the jet can be used to characterize the disk-jet connection. However, evidence of such link have been scarce and indirect. Mrk 421 is a BL Lac object which has a prominent jet pointed towards us and a weak disk emission, and it is assumed that most of its X-rays are generated in the jet. Hence, existence of the break in its X-ray PSD may indicate that changes in the accretion disk, which may be the source of the break timescale are translating into the jet, where the X-rays are produced. 
\end{abstract}


\section{Introduction}
Fluctuation in the emitted flux of radiation at multiple wave bands is a prominent characteristic of many active galactic nuclei (AGN). Consequently,  quantitative analysis of variability has often been used to probe the structure and physical mechanism in AGN. In some classes of AGN, e.g., Seyfert galaxies, X-rays are predominantly produced close to the black hole (BH) by inverse-Compton (IC) scattering of the optical/UV seed photons from the accretion disk by the so called ``corona", which is supposed to be a distribution of energetic electrons. Hence, the X-ray variability may be used to probe the properties of accretion on to the BH. Black hole X-ray binaries (BHXRBs) and Seyfert galaxies have been shown to have power-law X-ray power spectral density (PSD) with one or more ``breaks" \citep{utt02,mch06,cha09,cha11}, i.e., the slope steepens below a characteristic ``break-timescale" (T$_B$). In many cases the PSDs of the former are expressed as combinations of broad Lorentzians, cutoff powerlaws, and sharp peaks representing quasi-periodic oscillations. AGN PSDs may have intrinsic shapes that have similarly detailed feature, but that cannot be verified with currently available data. T$_B$ is proportional to the mass of the central BH and inversely related to the accretion rate \citep{mch06}. This may imply a universality in the accretion of matter on to BH of mass 10 to $10^8$ M$_{\sun}$. 

Blazars are a class of AGN with a prominent jet pointed close to our line of sight. Most of the observed X-ray emission in the blazars originates in the jet as opposed to the disk-corona region. That is primarily due to the amplification by a factor of 10 to 10$^4$ of jet emission due to relativistic beaming. PSD of multi-wavelength variability of blazars have shown scale-invariant behavior which could be fit well by a simple power-law \citep[e.g.,][]{cha08,nak13,shim13}. PSDs of 6-months-long radio, X-ray, GeV, and TeV light curves of Mrk 421 were determined by \citet{ale15}, \citet{goy17} studied the GeV, optical, and radio PSD of the BL Lac object PKS 0735+178, \citet{abe17} calculated the X-ray PSD of PKS 2155-304 from light curves covering a few days, \citet{cha12} analyzed 2-yr-long optical-near infrared (OIR) light curves of six blazars from the Yale/SMARTS blazar monitoring program, and \citet{bha16} analyzed the $\sim$110 hr nearly continuously sampled OIR variability of the blazar S5 0716+714. None of the above studies found a break in the PSD.

There have been some reports of breaks in the PSD of possible jet variability in the literature. However, it is not clear if those breaks are of similar nature to the break in the X-ray PSD of BHXRBs and Seyferts. \citet{moh15} found a bend in the 4.8 GHz - 36.8 GHz radio PSD at 0.52 $-$ 0.66 yr timescale. \citet{moh16} found a break at 6.2 hr timescale in the PSD of one out of six sections of a 1.6 yr long \textit{Kepler} light curve of the blazar W2R 1926+42. A break timescale of $\sim$2.7 yr was found in the PSD of the optical variability of the blazar PKS 2155-304 in its light curve covering variability at $\sim$20 days to $\sim$10 yr timescale \citep{kas11}. 
\citet{nak13} analyzed the \textit{Fermi}-LAT GeV light curves of 15 blazars. They found a break in the blazar 3C 454.3. 
\citet{mch08} studied the X-ray (2-10 keV) variability of the blazar 3C 273 and found a break in the PSD at $10^{-6.1^{+0.4}_{-0.4}}$ Hz (McHardy, I. private communication).
\citet{kat01} found a similar break in the X-ray PSD of 3C 273. \citet{shim13} analyzed 58-month-long hard X-ray (14-150 keV) light curves of 30 AGN and did not find any break in their PSD except in 3C 273. The break frequency they find ($10^{-6.8^{+0.3}_{-0.4}}$ Hz) is marginally consistent with that found by \citet{mch08}. However, it is unclear if the X-ray variability in 3C 273 is indeed due to the jet emission. Significant fraction of the X-rays in 3C 273 may be due to the IC scattering of the disk photons in the corona \citep[e.g.,][]{kal17}. Furthermore, OIR light curves of 3C 273 from the Yale-SMARTS blazar monitoring program from 2008-now exhibit a constant component, which is larger than the other blazars in the sample \citep{bon12}. This indicates that the relative fraction of the disk emission is larger in 3C 273 than that in the other blazars. Hence, the timescale found by \citet{mch08} may be directly related to the disk and not the jet emission. We note that \citet{mch08} found the break frequency to be $\sim 10^{-6}$ Hz. Therefore, if the PSD does not extend sufficiently around those ferquencies, i.e., say, from  $10^{-8}$ to $10^{-4}$ Hz, determining the break with confidence may not be possible in other sources.

\citet{kat01} analyzed X-ray light curves of three blazars from Advanced Satellite for Cosmology and Astrophysics (ASCA) including Mrk 421. They found possible breaks in the X-ray PSD of these blazars. However, due to the lack of long-term light curves at that time, the low-frequency part of the PSD did not extend below $2 \times 10^{-6}$ Hz, and hence the break frequency was not well-constrained. More recently, \citet{iso15} analyzed MAXI data to find a power-law-like PSD for the X-ray variability of Mrk 421 with a best-fit slope $1.60\pm0.25$ at the frequency range $10^{-8} - 2 \times 10^{-6}$ Hz. That slope is smaller than $2.14\pm0.06$ obtained by \citet{kat01} in the higher-frequency range ($\gtrsim 10^{−5}$ Hz). From that they infer the existence of a possible break and put a lower limit on the break frequency at $5 \times 10^{-6}$ Hz. 

In this paper, we analyze densely sampled light curves of the blazar Mrk 421 observed by \textit{AstroSat} during 2017 January along with long-term variability data from the \textit{Swift-XRT} and the \textit{Swift-BAT}. We construct the X-ray PSD of Mrk 421 at a much broader range of timescales than \citet{kat01} or \citet{iso15}, from $\sim$hr to years (frequency of $3.2 \times 10^{-9}$ to $1.3 \times 10^{-4}$ Hz), and carry out a comprehensive search for the presence of a break. In {\S}2 we describe the data reduction, we carry out the power spectral analyses in {\S}3, and finally we present the results and interpretation in {\S}4.

\section{Observations and Data Analysis}
\astrosat is India's first dedicated space telescope \citep{Rao2016}. It was launched in 2015 September. It can observe the sky simultaneously with all five of the instruments onboard at near and far UV, and soft and hard X-rays (up to 150 keV). We observed \source  with the Soft X-ray Telescope (SXT) and the Large Area X-ray Proportional Counter (LAXPC) instruments  onboard \astrosat during 2017 January 3 to 8 for a total of 400 ks. SXT and LAXPC observe at 0.3$-$8 keV and 3$-$80 keV energies, and their observing efficiencies are $\sim$25\% and $\sim$40\%, respectively. Hence, the total exposure time is 98 ks for SXT and 180 ks for LAXPC. 

\subsection{SXT}
SXT is a focusing telescope sensitive to 0.3$-$8.0 keV energies \citep{Singh2016SPIE, Singh2017JApA}. The Level-1 SXT data are first analyzed using the {\tt sxtpipeline} tool provided as a part of the recent SXT data analysis package ({\tt AS1SXTLevel2-1.3}). This pipeline generates the Level-2 cleaned event files for all orbits individually, which are then merged using the auxiliary tool {\tt SXTEVTMERGER} provided by the instrument team\footnote{http://astrosat-ssc.iucaa.in/?q=data\_and\_analysis}. The merger script takes care of the duplicate events and time-overlap, if any. The resulting event file is then used to generate the source spectrum, lightcurves and the image using the standard \textit{heasoft} tools. A circular area with a radius of 13$'$ centered at the source position is used to make the source region. Background counts for respective energies, derived by the instrument team using deep background observations across the sky, are used to correct the observed lightcurves. This is performed to avoid introducing extra noise in the source light curve because of the extremely low statistics in the background light curves made from the thin 15$'-$18$'$ annulus available around the source region. While it is possible that a small amount of artificial variation may be introduced if a simultaneous background is not used we note that no high solar activity was reported close to our observing period and the variability at the soft and hard bands of the SXT and LAXPC are strongly correlated, which indicate that the observed variability is not significantly affected by the background variations, if at all. 
\begin{figure*}
\centering
\includegraphics[scale=0.7,angle=270]{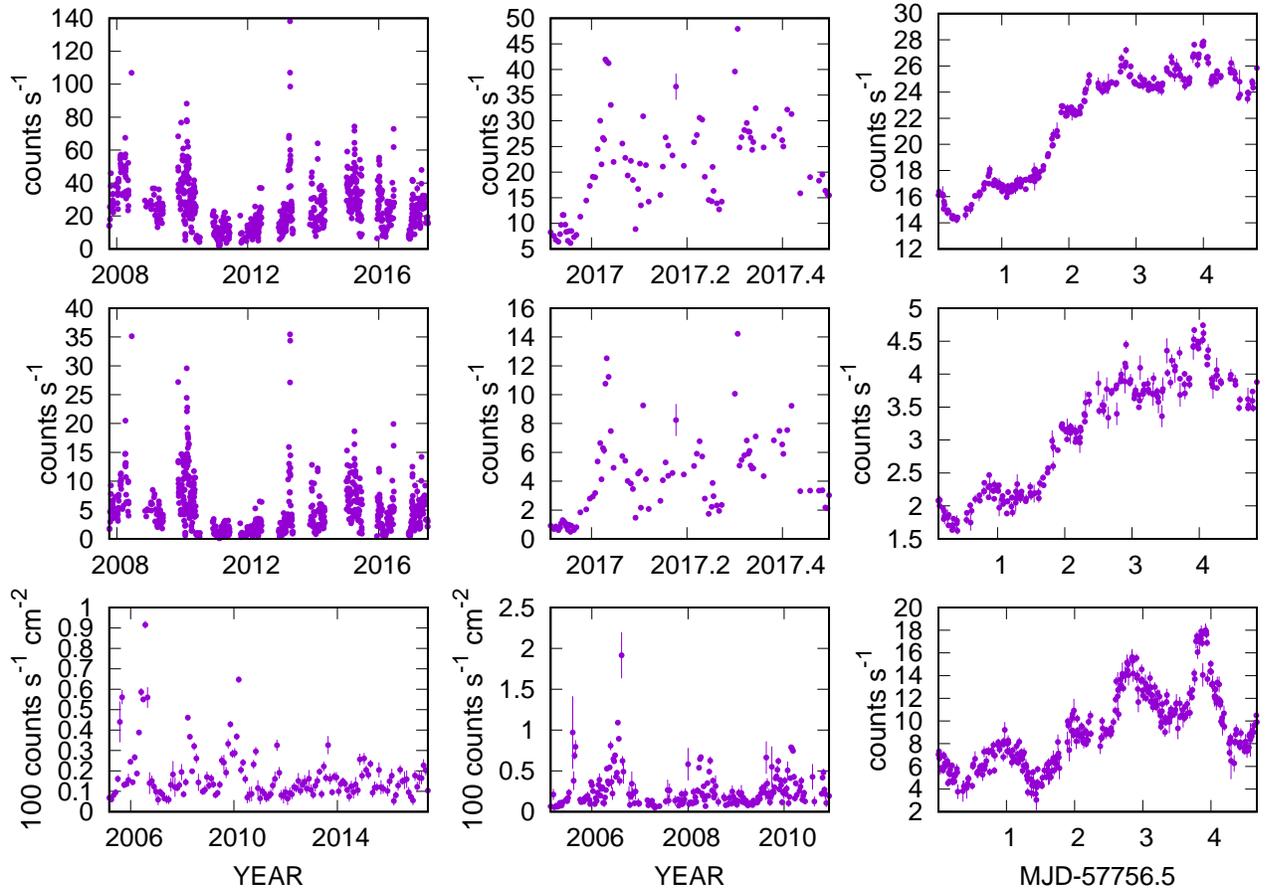}
\caption{The X-ray light curves of Mrk 421 at various timescales and energy bands. The top row shows the low-energy X-ray light curves (left and center panels: 0.3$-$2 keV from the \textit{Swift-XRT}; right panel 0.7$-$2 keV from the \astrosat SXT), the middle row exhibits the medium-energy X-ray light curves (left and center panels: 2$-$8 keV from \textit{Swift-XRT}; right panel 2$-$8 keV from the \astrosat SXT) while the bottom row shows the high-energy X-ray light curves (left and center panels: 15$-$50 keV from \textit{Swift-BAT}; right panel 15$-$50 keV from the \astrosat LAXPC).}
\label{lcall}
\end{figure*}
\subsection{LAXPC}
The LAXPC consists of a set of three identical units of proportional counters making the effective detector area $\sim$6000 cm$^2$ and nearly co-aligned with the satellite axis \citep{Antia2017}. We use the {\tt laxpc\_soft} package, available at the \astrosat Science Support Cell (ASSC) website\footnote{http://astrosat-ssc.iucaa.in/?q=laxpcData}, to analyze the LAXPC Level-1 data and use the recommended analysis procedures \citep{Antia2017, Yadav2016}. The list of Level-1 data tree in the appropriate format including all data and attitude files, as suggested in the help document, is used as input along with other information such as energy range and binsize entered interactively from the command line. The South Atlantic Anomaly, Bright-Earth and the other attitude related corrections are made using the default parameters given by the instrument team. This package generates the Level-2 standard products, namely, spectra, and lightcurves for the source and the background. Another script, namely, {\tt backshift.f}, provided as part of the package, is used to correct for the drift in the gain of the detectors with time, if any. To optimize between inclusion of photons of energy up to 50 keV and minimising the cosmic ray background, we include in our analyses count rates from the top 2 layers.

\subsection{Other Archival Data}
We use the archival X-ray light curves from the \textit{Swift-XRT} Monitoring Program\footnote{http://www.swift.psu.edu/monitoring/}, which provides near real-time results from the \textit{Swift} observations of the \textit{Fermi-LAT} ``sources of interest" and flaring sources \citep{str13}. Furthermore, we use 15-50 keV X-ray light curves from the \textit{Swift-BAT} Hard X-ray Transient Monitor\footnote{https://swift.gsfc.nasa.gov/results/transients/} \citep{kri13}. The X-ray light curves used in this paper are shown in Figure \ref{lcall}. 

\section{Analyses and Results}
Raw PSD calculated from a light curve includes the effects of the generally irregular temporal sampling pattern of the observations in addition to the intrinsic variation of the object. Furthermore, PSD suffers from the distorting effects of the finite length (``red noise leak") and the discontinuous sampling (``aliasing") of the light curves. 
To avoid the above difficulties we use a Monte-Carlo type approach similar to the ``Power Spectral Response Method," \citep[PSRESP;][]{don92,utt02, cha08} to calculate the PSD. We simulate many sets of light curves using a range of model PSDs and including the distortions as described above. The underlying PSD is determined through the comparison of the distribution of the PSDs of the simulated light curves to that of the observation. By including the imperfections present in the observation in the simulated light curves, a physically meaningful comparison of the observed PSD with the distribution of the simulated PSDs is possible. Our method gives a best-fit PSD model and a so called ``success fraction'' $F_{\rm succ}$ (the fraction of the simulated light curves that successfully represent the observed light curve) that indicates the goodness of fit of the model. 

To generate the artificial light curves above we follow \citet{tim95}, which is appropriate only for simulating Gaussian distributed light curves. However, it has been shown \citep{utt01} that long-term light curves of blazars can have non-Gaussian distribution, e.g., log-normal. \citet{ema13} have put forward a method to generate artificial light curves which can exactly mimic the distribution of the observed light curves or any other model. Implementing that method is beyond the scope of this work. We assume that changes in our results due to using Gaussian distributed artificial light curves, if any, will be within the uncertainties that we find.

Initially, we attempt to fit a simple power-law model to the X-ray PSD, but find that the best-fit $F_{\rm succ}$ is low ($0.2$). This implies that a simple power law is not the best model for this PSD. We therefore fit a bending power-law model (broken power-law with a smooth break) to the X-ray PSD \citep[][]{mch04}:
\begin{equation}
P(\nu)=A\nu^{\alpha_L}[1+(\frac{\nu}{\nu_{B}})^{(\alpha_{L}-\alpha_{H})}]^{-1}.
\end{equation}
Here, $A$ is a normalization constant, $\nu_{B}$ is the break frequency, and $\alpha_H$ and $\alpha_L$ are the slopes of the power laws above and below the break frequency, respectively. During the fitting, we varied the break frequency $\nu_{\rm B}$ from $10^{-9}$ to $10^{-4}$ Hz in steps of $10^{0.05}$, $\alpha_{\rm H}$ from $-2.0$ to $-3.0$ in steps of 0.1, and $\alpha_{\rm L}$ from $-0.6$ to $-2.0$ in steps of 0.1. These ranges include the values of $\alpha$ found in the light curves of BHXRBs, for which $\alpha_{\rm L} \approx -1$ and  $\alpha_{\rm H}$ is between $-2$ and $-3$ \citep[e.g.,][]{rem06}. This procedure yields a much higher success fraction than the simple power-law model. PSDs at the three energy bands are shown in Figure \ref{psdall}. The best-fit parameters based on the model with the highest success fraction are given in Table 1. {Variation of success fraction with changes in any two of $\nu_{B}$, $\alpha_H$, and $\alpha_L$ while the other parameter is kept fixed at their average value are shown in Figure \ref{contour}. The 1-$\sigma$ uncertainties given in Table 1 are found from Figure \ref{contour}.

Using $M_{\rm BH} = 1.9 \times10^8$ M$_{\sun}$ \citep{woo02}, $L_{\rm bol}$ = $\rm 10^{-3}L_{edd}$ $=2.5 \times 10^{43}$ erg\,s$^{-1}$, and the best-fit values and uncertainties in the $\rm T_B$---$\rm M_{\rm BH}$---$\rm L_{\rm bol}$ relation proposed by \citet{mch06}, expected break frequency ($\rm T_B^{exp}$) is $10^{-7.9\pm0.8}$ Hz. We note that the uncertainty in this value of $\nu_{B}$ represents that in the $\rm T_B$---$\rm M_{\rm BH}$---$\rm L_{\rm bol}$ relation. However, the BH mass is estimated from the stellar velocity dispersion and may contain uncertainties of $\sim$1 order of magnitude. The value of $L_{\rm bol}$ is an estimate which corresponds to an accretion rate $\rm \sim 10^{-3} ~L_{edd}$. Therefore, the actual uncertainty in the expected value of $\nu_{B}$ is difficult to quantify. The value of the break frequency we find here is $10^{-5.7\pm1.0}$ Hz. The bulk Lorentz factor of the high-energy emission region in the jet of Mrk 421 has been estimated to be $\sim$20 by, e.g., \citet{ghi10,lic12}. If $T_B$ is indeed shorter than $\rm T_B^{exp}$ by a factor of $\sim$20, it may indicate that the break timescale is compressed due to relativistic beaming in the jet. It may also imply that the break we find in Mrk 421 does not follow the $\rm T_B$---$\rm M_{\rm BH}$---$\rm L_{\rm bol}$ relation and is due to some other physical process, possibly in the jet.
\begin{table}
\begin{center}
\caption{Parameters of the Best-Fit Power Spectral Models.\label{psd_table}}
\begin{tabular}{cccc}\\
\tableline\tableline
Energy Band & $\alpha_L$ & $\alpha_H$ & $\nu_B$ (Hz) \\
\tableline
0.3$-$2 keV		&	$-1.4^{+0.6}_{-0.4}$	&	$-2.5^{+0.5}_{-0.5}$	&	$10^{-5.7 \pm 1.0}$	\\
2.0$-$10 keV	&	$-1.2^{+0.5}_{-0.5}$	&	$-2.4^{+0.5}_{-0.5}$	&	$10^{-6.0 \pm 1.0}$	\\
15$-$50 keV	&	$-1.1^{+0.4}_{-0.4}$	&	$-2.5^{+0.5}_{-0.5}$	&	$10^{-5.5 \pm 0.6}$	\\
\tableline 
\end{tabular}
\end{center}
\end{table}

\section{Discussion and Conclusion}
Presence of a confirmed break in the X-ray PSD of Mrk 421 implies that there is a characteristic timescale in its X-ray variability. Mrk 421 is a high-frequency-peaked BL Lac object (HBL). One of the defining properties of such objects is the lack of emission lines and weak disk emission. Hence, most of the observed X-rays are produced in the jet \citep[e.g.,][]{abd11}. The physical origin of the X-ray variability may be in the jet itself or it may originate in the disk and translate into the jet.
\begin{figure}
\epsscale{1.2}
\plotone{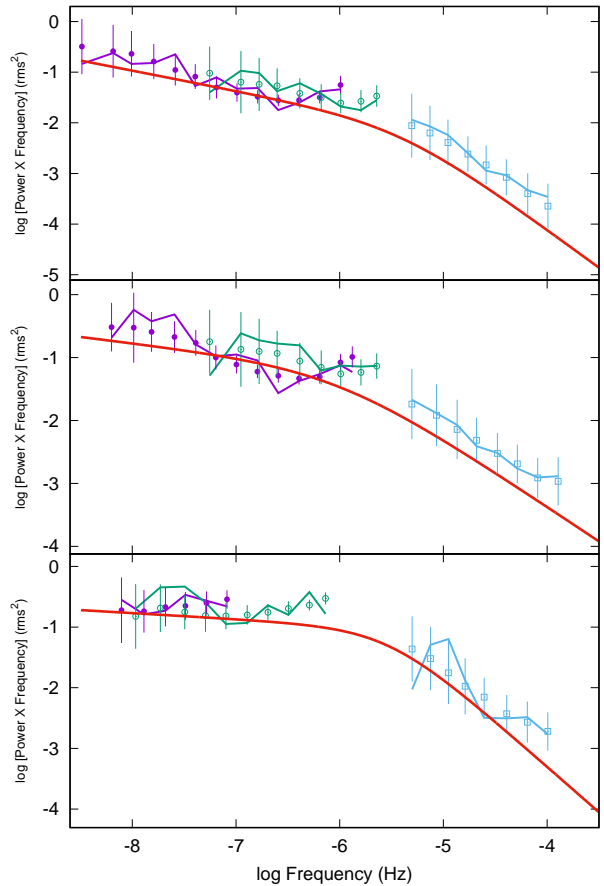}
\caption{PSD of the 0.3$-$2 keV X-ray variability at low, medium, and high frequencies are given by the purple, green, and cyan jagged lines, respectively, in the top panel while the underlying best-fit bending power-law model is given by the thicker red solid bent line. Points with error bars (filled circles, open circles, and open squares for low, medium, and high frequency range, respectively) correspond to the mean value of the PSD simulated from the underlying power-law model (see text). The error bars are the standard deviations of the distribution of the simulated PSDs. The middle and the bottom panels show the same for the 2$-$10 keV and the 15$-$50 keV X-rays. Here, the three panels represent the three rows while the low, medium, and high frequencies correspond to the three columns in Fig. \ref{lcall}.}
\label{psdall}
\end{figure}
\begin{figure}
\epsscale{1.2}
\plotone{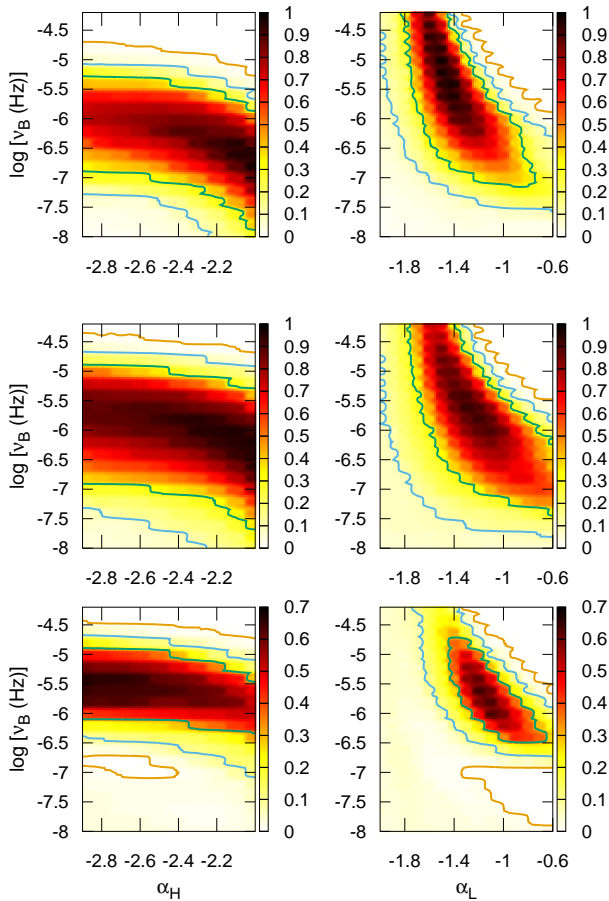}
\caption{The top panels show the variation of success fraction with the parameters of the bending power-law model in the case of the 0.3$-$2 keV X-ray PSD. The contours show 99 (orange), 90 (cyan), and 68 (green) percent confidence intervals. The middle and the bottom panels show the same for 2$-$10 keV and 15$-$50 keV X-rays.}
\label{contour}
\end{figure}

\citet{fin14} developed a theoretical model, which produces the PSD of blazar variability associated with nonthermal emission processes in the jet. They associate bends in their theoretical PSD with characteristic timescales of the emission process such as the radiation cooling, light-crossing or escape timescale. However, the timescale we have found here ($\sim$few to tens of days) is much longer than even the upper limit of the range of the shortest variability timescales (time interval during which the flux value changes by a factor of 2) observed in the blazars ($\lesssim$hr). Hence, that cannot be interpreted as the crossing timescale as in that case variability at shorter timescales will be smoothed out and will not be observed. The cooling timescale of particles emitting X-rays in blazar jets is shorter ($\lesssim$1 day) than the PSD break timescale found here \citep[e.g.,][]{sai13}. However, they find the frequency corresponding to the escape timescale in the observer's frame to be $\sim 2 \times 10^{-5}$ Hz, which is marginally consistent with the break frequency we observe within the uncertainties. In that case, $t_{esc}$ at the observer's frame is $\frac{0.5 day}{2\pi} \simeq 2$ hr and at the jet frame $\sim$20 hr. Furthermore, in \citet{fin14}, the difference in slope between the frequencies below and above the break is $\sim$2.0, and the low-frequency slope of the PSD is zero if the parameter ``$a$" related to injection is set to zero. That is not consistent with observations. However, the low-frequency slope $\sim-1$ if $a$ is set to non-zero values. That is consistent with our results.
 
In another theoretical model put forward by \citet{che16} the difference between the low-frequency and high-frequency slopes of the PSD $\sim$2.0, and the low-frequency slope is zero. These are not supported by our results. However, in their model, the relevant physical timescales in the system are cooling timescale, crossing timescale and the so called ``acceleration decay" timescale ($\tau_{\rm decay}$). They find $\tau_{\rm decay} \simeq$ 3 days in the jet frame (corresponding to a frequency $\sim 4 \times 10^{-6}$ Hz) and hence $\sim$ 0.3 day in the observer's frame, which is marginally consistent with the $T_{\rm B}$ we observe within the uncertainties.

As stated by \citet{che16}, the above timescales and slopes may vary depending on the various parameters of their model. A small and realistic change in the parameters may make the above match with observed results even better. Consequently, stronger constraints on their models may be imposed in order to match the relevant timescales exactly with the results obtained here for Mrk 421.

Modeling AGN variability at optical, X-ray and $\gamma$-ray bands with Ornstein-Uhlenbeck (OU) and mixed OU processes have been carried out extensively \citep[e.g.,][]{kel11}. They find characteristic timescales in most light curves, unlike the works that model the PSD. It is unclear if the PSD break timescale and the timescales found by the OU and mixed OU modeling are of the same physical origin \citep{fin14}. We note that if the low-frequency slope is assumed to be zero in our method in accordance with OU and mixed OU modeling, the resultant best-fit break timescale may be different.

 The above discussion implies that the break timescale found in the X-ray PSD of Mrk 421 is consistent with what \citet{kat01} and \citet{iso15} tentatively obtained. It is similar to the break timescale found in the X-ray PSD of BHXRBs and Seyfert galaxies, and probably distinct from the characteristic timescales which is a parameter in the OU and mixed OU models used to fit AGN light curves. As BL Lac objects are believed to contain a weak accretion disk, the timescale definitely exists in the jet emission. However, since the PSD break timescale originates in the disk-corona region of BHXRBs and Seyfert galaxies, it is possible that this is an evidence of translation of disk variability to that in the jet emission. We note that we have assumed the PSD of Mrk 421 to be stationary, i.e., its slope and normalization do not change with time. Consistency of the slope we find with that of \citet{kat01} using data from $\sim$20 yr ago supports our assumption. To test the effect of a variation in the normalization of the PSD, we artificially decrease and increase the normalization of the 0.3$-$2 keV SXT PSD by 50\%. The resultant best-fit break frequency changes from its original value of $4 \times 10^{-6}$ Hz to $3 \times 10^{-6}$ Hz and $6 \times 10^{-6}$, respectively, which are well within the uncertainties. Hence, we expect the break timescale we find here to be robust. However, we do plan to test the assumption of stationarity with a large number of X-ray light curves of Mrk 421 in a future work.
 
It is believed that the observed X-ray variability from the accretion disk-corona region in the BHXRBs and Seyfert galaxies is due to a physical scenario similar to that described by \citet{lyu97}. In that model, fluctuations propagate from the outer radii of the accretion disk towards the center and couple with those originating in the inner radii. These fluctuations in the disk plasma may translate into the jet and manifest themselves as the emission variability in the jet \citep{iso15}. In that case, characteristic timescales present in the PSD of the disk-corona variability may be present in that of the jet variability. The timescale may be modified due to relativistic beaming or stay in tact \citep{mch08}. 

An important consequence of the above model is a multiplicative contribution of the underlying processes to the observed flux, rather than an additive one. This predicts that the observed flux distribution should be lognormal, as opposed to Gaussian. While such lognormal distributions are regularly found in the light curves of BHXRBs \citep{utt01} and Seyfert galaxies \citep{vau03}, they are also being seen in the long-term flux distribution of blazars at different wavelengths \citep[e.g.,][]{gie09}, and specially for Mrk 421 at X-ray and GeV energies \citep{sin16}. Such observations have independently been proposed to be indicative of the variability imprint of the accretion disk onto the jet \citep{mch08}.

The break timescale ($T_B$) that we find here as well as those in BHXRBs and Seyfert galaxies is much longer than the dynamical timescale $(R^3/GM)^{1/2}$ at the inner edge (at radius $R$) of the accretion disk. Hence, other physical origin for $T_B$ such as a viscous timescale or characteristic timescale for large scale alignment of poloidal magnetic field in the inner accretion disk from random fluctuations have been proposed \citep{kin04,cha09}.

We do not find a significant variation in the break timescale as the photon energy changes from 0.3$-$2 keV to 15$-$50 keV. This is is not consistent with the trend observed in the Seyfert galaxies that the slope is flatter at higher energies \citep[e.g.,][]{gon12}. However, this is consistent with the suggestion that the process causing the variation does not occur within the jet \citep{mch08}. Harder X-rays may be produced by higher energy particles (e.g., electrons), which have a shorter lifetime. Therefore, the timescale related to the harder X-rays may be shorter than that related to the softer X-rays. We note that given the relatively large uncertainty in the break timescale such difference may exist and yet not be detected. However, if variation from outside the jet, e.g., the accretion disc, modulate the jet emission or produce moving shocks in the jet, which eventually cause the variation in the jet X-rays, then the characteristic timescale found may be the same for all bands, as seen here.

\section{Acknowledgements}
This publication has made use of data from the Astrosat mission of the Indian Space Research Organisation (ISRO), archived at the Indian Space Science Data Centre (ISSDC). RC received support from the UGC start-up grant and Presidency University FRPDF. ARC received support from the DST INSPIRE fellowship. ARC thanks Gulab Dewangan for advice related to SXT data analyses. AS will like to thank the LAXPC instrument team for useful discussion. RC thanks IUCAA for their hospitality and usage of their facilities during his stay in 2017 summer as part of the university associateship program.\\

\end{document}